\documentclass[11pt, graphicx]{article}
\usepackage{delarray}
\usepackage{epsfig}         

\begin{document} 

\begin{center}
\Large\textbf {{Numerical Simulations Of Particle Acceleration In Relativistic 
Shocks With Application To AGN Central Engines\footnote{Results presented in the \textit{Particles Fields and Radio Galaxies Workshop}, August 2000, Oxford UK }}}
\end{center}

\begin{center}
\renewcommand{\thefootnote}{\fnsymbol{footnote}} 
\Large {John Quenby \& Athina Meli }
\end{center}
\begin{center}
{\small {\textit {Astrophysics Group, Blackett Laboratory}}}
{\small {\textit{Imperial College of Science,Technology and Medicine, London, 
UK}}}
\end{center}

\begin{abstract}
{\small\textbf {Numerical modelling is performed for extreme relativistic 
parallel shocks
with upstream Lorentz factor $\Gamma=50$. Assuming the scattering is either 
large angle or over pitch angles $\sim \Gamma^{-1}$,
spectral flattening and shock aaceleration speed-up is found. The energy gain
for the first shock cycle is $\sim \Gamma^{2}$. The likely output from relativistic shocks 
due to the infall from the accretion
disk to the AGN black hole is computed. Neutrinos from proton-gamma
interactions may be detectable
with planned neutrino telescopes}}. 
\end{abstract}

\section{\Large{Diffusive Shock Acceleration}}
\large {Relativistic plasma flow velocities are found in many 
astrophysical objects such as in AGN Jets and their Central Engines where transient shocks are thought to be formed and diffusive particle acceleration could potentially occur. The gamma 
flows at these sites are approximatelly  5-10 Lorentz Factors. Gamma Ray Burst(GRB) plasma flows also appear 
to have ultra-relativistic velocities where $\gamma \sim 100-1000$. 
Particularly Waxman (1995) and Vietri (1995,1997 and 1998) have 
noted (for the case of CRBs) the fact that 
diffusive acceleration at non-relativistic shocks is not fast enough to produce 
the Highest Energy Cosmic Rays (HECR) observed. Vietri (1995) especially, pointed out that 
simulations of Quenby $\&$ Lieu (1989) were the first to  recognize a considerable 
enhancement in the acceleration rate of a  factor$\sim$13, for relativistic shocks 
of $\gamma$$\sim$3.

Our current investigation for highly relativistic plasma flows,  shows a dramatic acceleration 
rate enhancement, and since the Lorentz  boosting scales
as $\gamma^{2}$ (especially in the first cycle)
this effect is important for the Vietri's HECR theoretical predictions. 
It is known that the  chief characteristics of diffusive 
shock acceleration operating in one
dimensional astrophysical non-relativistic shocks are the $\gamma^{-2}$ 
differential spectra for 
particle Lorentz factor $\gamma$ when the shock compression ratio is 4 and the
analytically derived acceleration time constant is given by, 

\begin{equation}
 t_{acc}=\frac{c}{u_{1}-u_{2}}\left[\frac{\lambda_{1}}{u_{1}}+\frac{\lambda_{2}}{_{2}}\right]
\end{equation}

where, $\lambda$ is mean free path, u is flow speed and 1 and 2 refer to upstream 
and downstream. 
For relativistic flows (eg. AGN jets, $\Gamma \sim 5-10$ and GRB
fire-balls $\Gamma \sim 100-1000$) the fractional energy increase for a single
shock crossing is $\sim (1-V^{2}/c^{2})^{-0.5}$ where V is the relativistic
difference in flow velocities across the shock,
but the analytical estimate of cycle time does not change from the
non-relativistic case and so we expect a speed-up of acceleration relative 
to the non-relativistic
estimate. However, because 
of the large anisotropies due to the relativistic flows, a computational 
approach and careful simulations are necessary
in the high $\gamma$ regime. Several studies have been made by few authors, 
but still, there is a controversy about the shape of the spectrum, the rate of 
energy  gain  and the time constant for acceleration. However, they  are crucial to 
models of AGN Jet and Central Engine cosmic ray 
acceleration as well as the GRB particle production and acceleration. In this 
paper  we present a computational study of relativistic diffusive shock acceleration for up to $\gamma$$\sim50$ and try to investigate 
the acceleration behaviour at these highly relativistic regimes. 

\subsection{\large{Numerical Simulations }}
\begin{figure}
\epsfig{figure=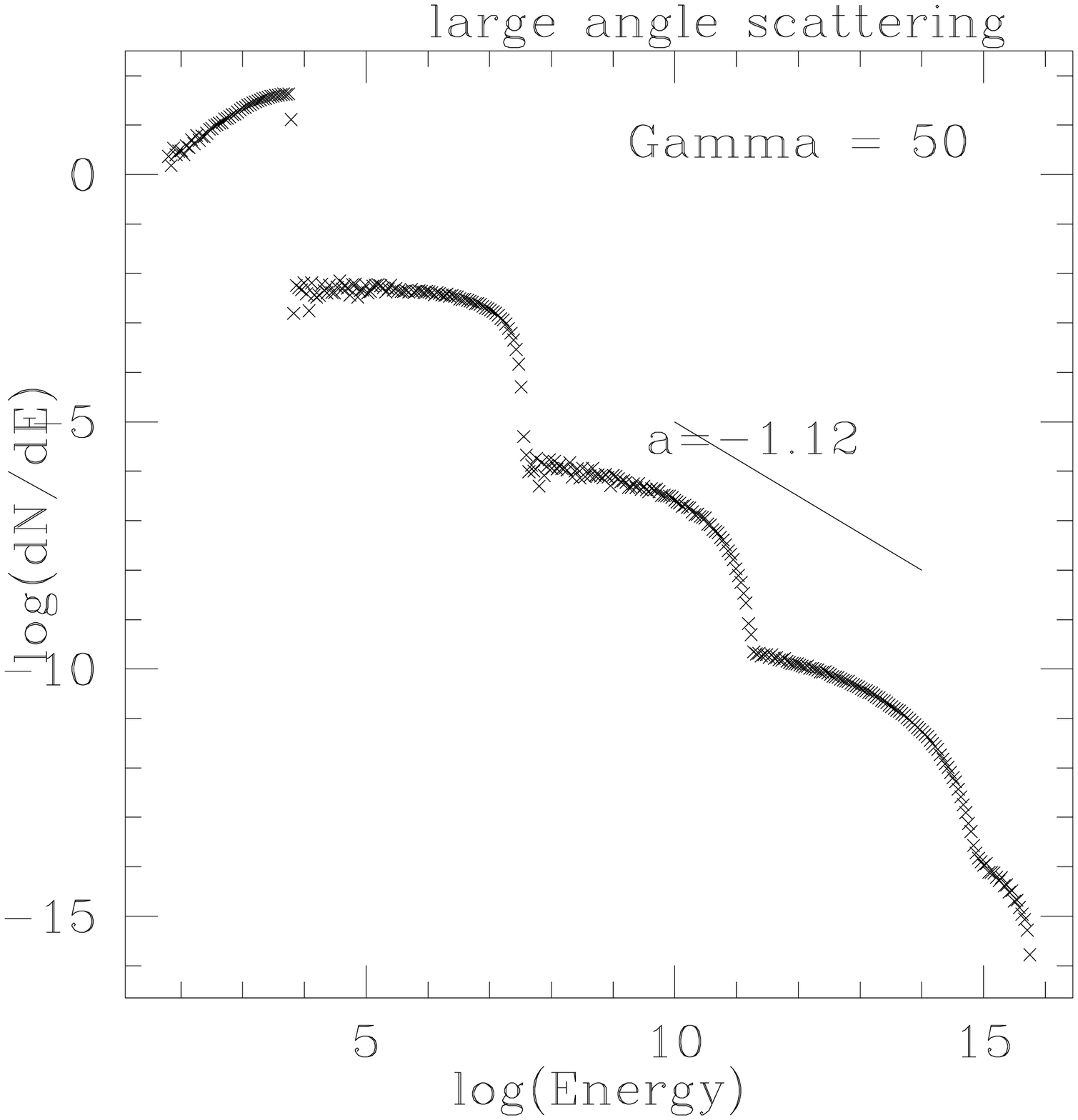,width=70mm}
\epsfig{figure=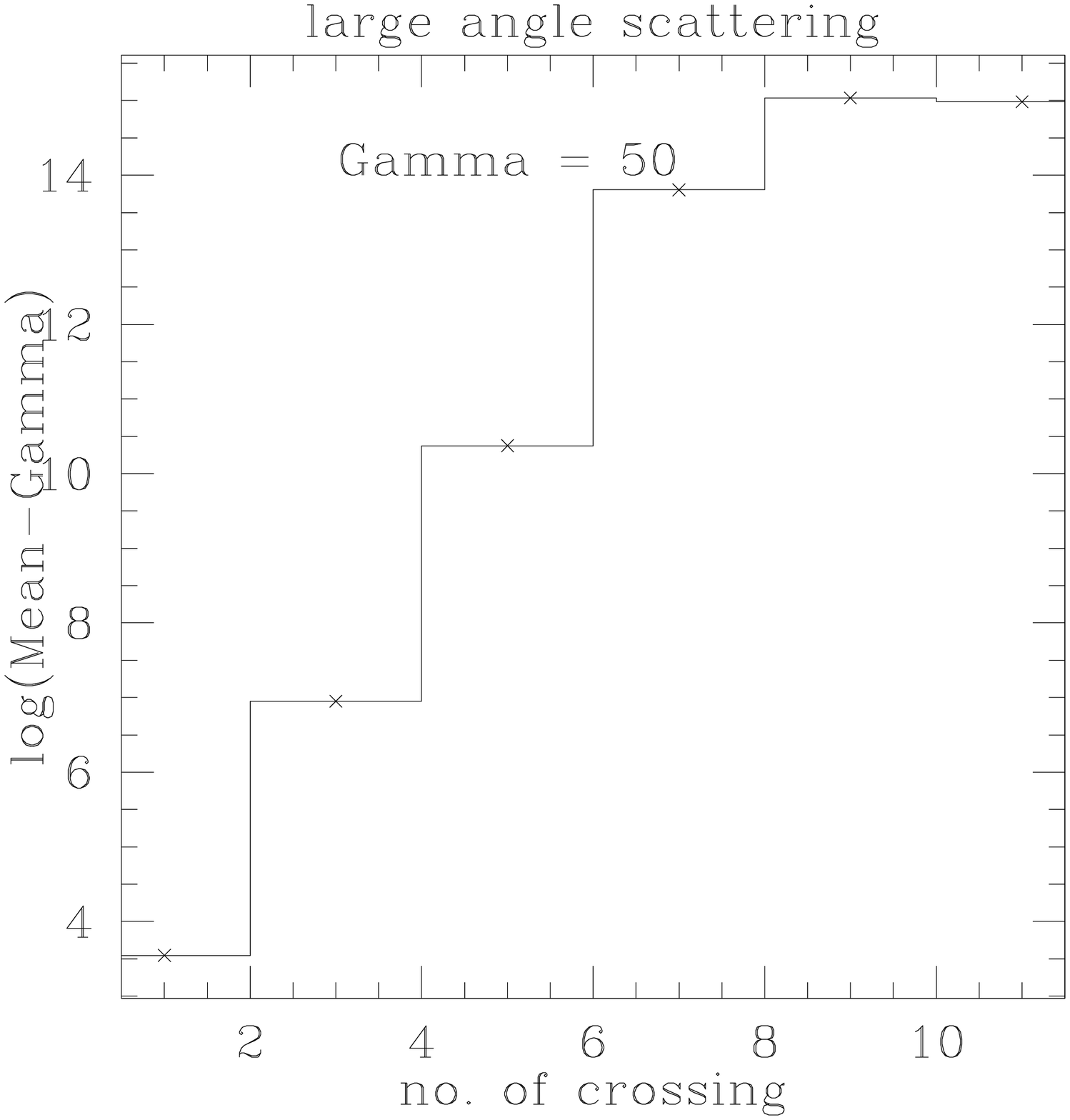,width=70mm}
\caption{Energy Spectrum (L), Mean Gamma Gain versus Crossing (R)}
\end{figure}

In these simulations we study the acceleration of particles by a numerical 
Monte Carlo approach. A guiding centre approximation 
is used for the particle's propagation in a 1D geometry 
and the shocks assumed to be "parallel", either because that is the field
configuration
or turbulence removes "reflection" at the interface. We keep a compression
ratio of 4 in order to be able
to be able to compare directly with non-relativistic situations
and consider large angle isotropic scattering in the plasma rest
frames according to,
\begin {equation}
Prob(z) \sim exp(-z/\lambda|cos\theta|)
\end{equation}
for parallel mfp $\lambda$,
proportional to rigidity and pitch angle $\theta$. Gallant \& Achterberg 
(2000) distinguish a small angle 
 scattering regime where $\theta \sim \Gamma^{-1}$ for scattering models where 
the field is either uniform or consists of cells of randomly orientated 
uniform field. For the time being we are not in this limit, but consider the cases
 where the particle coming upstream from the shock moves a distance $\sim
\lambda $ with only a small net
total deflection, during which time it must be travelling close to the shock
normal, 
otherwise it could not penetrate upstream against the relativistic convective
flow.
Then a turbulent field or flow discontinuity is encountered and deflection
takes place in a time $\sim r_{g}/c$ where $r_{g}$ is the gyroradius. 
Provided $\lambda > r_{g}\Gamma$, significant
deflection may take place before the particle is swept back to the shock. 
Lorentz transforms are used between frames on shock passage and the results
recorded just downstream in the shock frame. In this numerical investigation and in order to keep the statistics of the results as reliable as possible, a particle 
'splitting' technique is used. The particles injection is upstream, 
where approximatelly $10^{6}$ particles have been used for the completion 
of the code runnings.

Results for isotropic scattering with an upstream $\Gamma \sim 50$
are shown in figure 1(L) where the energy scale is measured by the particle
\begin{figure}  
\begin{minipage}{170mm}
\epsfig{figure=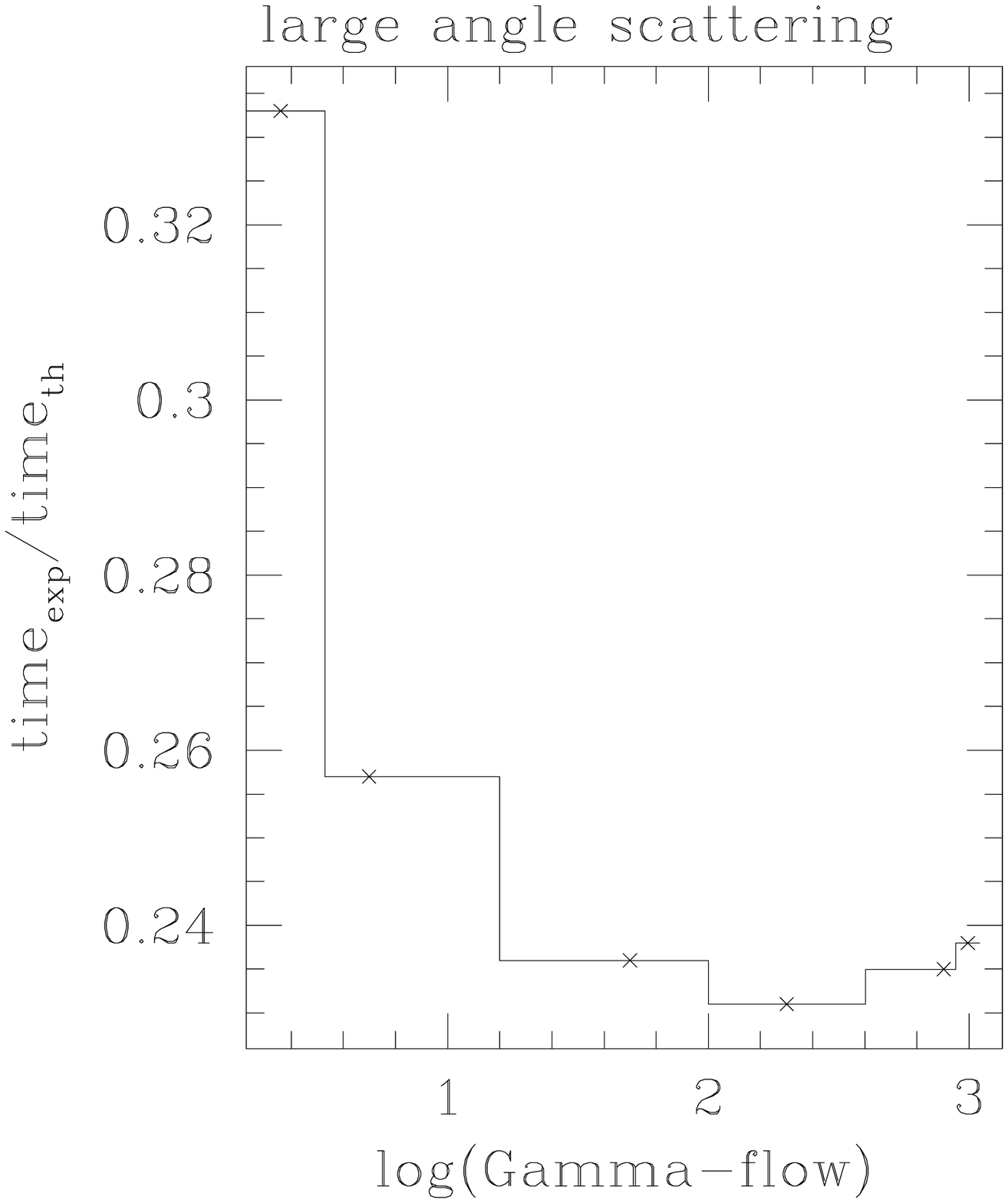,width=70mm}
\epsfig{figure=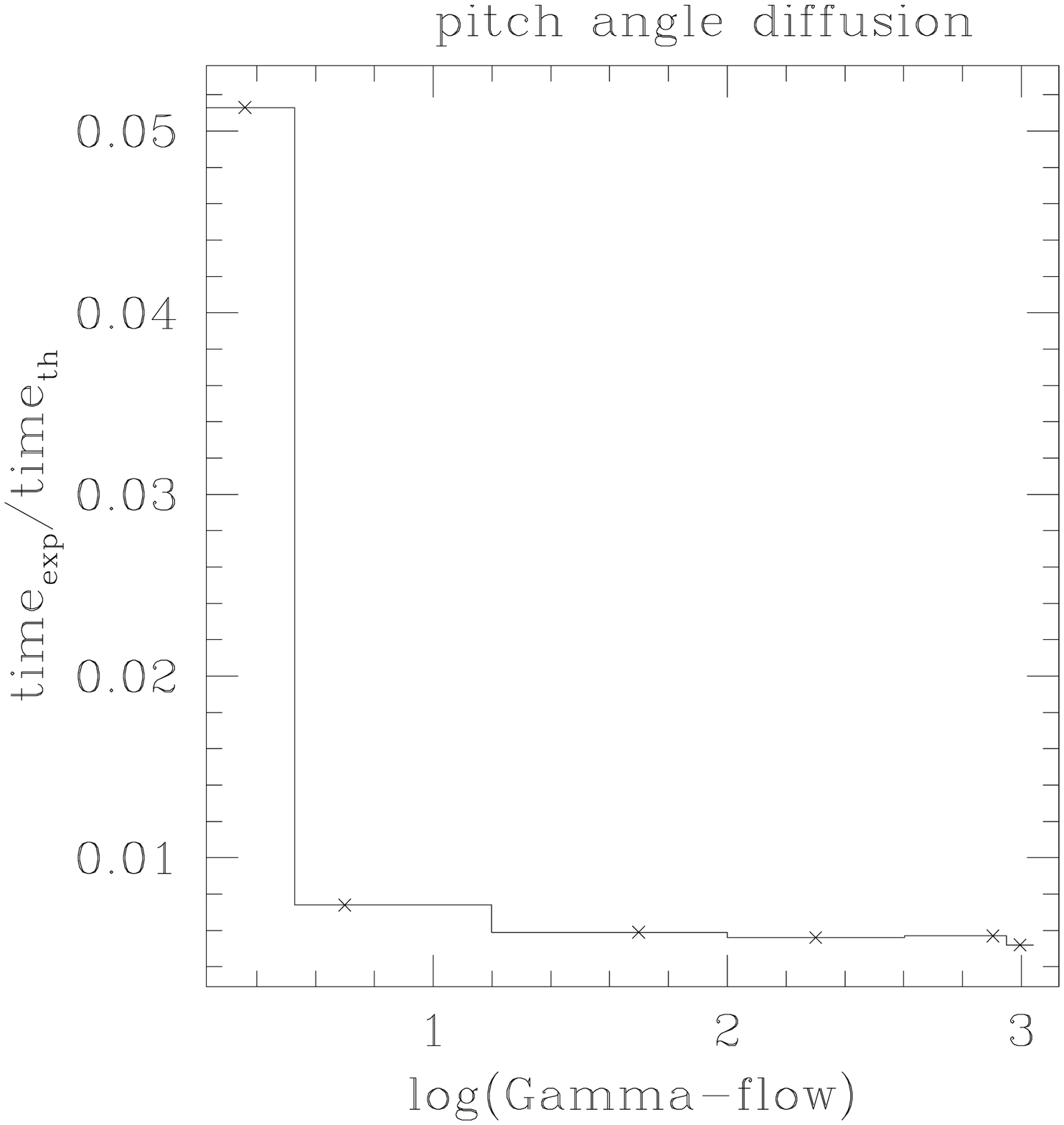,width=70mm}
\caption{Acceleration Speed-Up, Isotropic Scatter(L), Pitch-Scatter(R)}
\end{minipage}
\end{figure}
Lorentz factor, together with the best fitting power law. Clearly other factors,  especially trapping, 
limit the actual high energy extent of the spectrum. 
The plateau-like structure 
corresponds to successive cycles of acceleration. Our calculations show that 
roughly $80\%$ of all
particles are lost downstream each cycle, because of the high plasma flows. 
The mean energy gain per cycle
on cycle 1 for isotropic scattering is 80 and on cycle 3 is 3200, to be
compared with the square of the upstream flow gamma 
factor $\Gamma^{2}_{up}$=2500. This is seen in figure 1(R) where injection is
at $\gamma=50$.
Figure 2(L) shows the ratio of the computational to the non-relativistic,
analytical acceleration time constants as a function of $\Gamma_{up}$, for isotropic
scattering and figure 2(R) for small-angle scattering. For small-angle scattering, 
the mean free path to scatter $\sim \pi/2$ 
is used in the analytical expression. As it may be seen, a substantial 
'speed-up' is noticed in both cases and along with our  results of the very large energy boosting in consecutive shock cycles, which establishes the \textit{allowed} upper energy limit in GRB fireballs, Vietri's theory is been supported regarding the predictions of the origin of Ultra High Energy Cosmic 
Rays from GRB fireball sites.

\section{\Large{AGN Accretion Shock Model}}

A preliminary application of the speed up and spectral hardening confirmed above has 
already been given by Battersby et al (1995). 
Central accretion shocks for infall to AGN black holes must be sporadic 
because stable shocks occur at a given position for a very specific 
plasma angular momentum and internal energy (Molteni, Lanzafame \& Chakrabarti
1994). We expect quasi-perpendicular shocks to form out to 100 Schwarzchild 
radii because quasi-keplerian plasma motion should be taken into account
but is neglected in Battersby et al (1995). We estimate the lifetimes as given by free-fall 
and get 'de Hoffmann-Teller' frame speeds $\geq 0.5c$  
Using the above calculations of speed-up and spectral slope and allowing a
realistic increase in
the  mean free path above cyclotron radius yields the acceleration time and
spectrum.
The input particle total energy and magnetic field is estimated assuming
$L_{X-ray}=0.05L_{Edd}$ is in equipartition with magnetic energy and
particle energy, $B \propto R^{-1}$ and the Stecker-Salamon photon spectrum is
used.
Proton scattering in the field and convection into black hole is followed
for radial infall and quasi-keplerian plasma motion.
The secondary cascade following initial acceleration and propagation is
followed. High energy protons cool
preferentially on photons
while low energy protons cool preferentially on background protons.
pion decay into $\gamma, \nu$ and $ e^{\pm}$ components
 and also a $\gamma$ cascade, synchrotron photon production and
$\gamma-\gamma$ interactions result.
Battersby et al (1995) concentrate on neutrino
production from p gamma interactions. These interactions are more
important than pp at high energies. They find that a 1 $km^{2}$ 
high energy neutrino telescope is really required to have a good 
chance of detecting fluxes from a source like 3C273 or the background from
a large number of sources.

\section{Conclusions}

A Monte Carlo numerical investigation has been reported on the 
test particle limit of parallel diffusive shock acceleration. Very high gamma flow 
astrophysical plasmas have been used  up to $\gamma_{up}$ 
$\sim50$ which are relevant to AGN Jet and Central Engines shocks as well as the GRB ultra-relativistic enviromments. A dramatic  $\gamma^{2}$ (especially for the first cycle) 
energy boosting has been observed and a significant acceleration rate efficiency has been 
reported. 
Especially, both these results support (and been supported by) Vietri's and 
Waxman's models and theoretical predictions on the problem of HECR origin from GRBs. This crucial issue though,  of ultra-relativistic diffusive shock acceleration, needs further investigation as many other parameters are needed to be included within the simulation codes in order for the model and the results to be as realistic as possible in connection with the observations and the theoretical predictions. 

Much of the secondary output from the AGN originates from sporadic shocks
near 100 Schwarzschild radii. pp interactions are less important than 
p$\gamma$
in neutrino production except at low energy.
$\gamma$s from synchrotron dominate $\gamma$s from $\pi$s up to 1 TeV,
but the GeV to TeV $\gamma$ flux is a factor 100 less than observation of 3C273
output, suggesting some "Beaming" or amplification mechanism is required if 
central accretion shocks are 
the main source. We expect significant fluxes from $<100 R_{Sch}$ if the short
period fluctuations
observed are related to black-hole physics. 
The $100 R_{Sch}$ proton output $\leq0.1$ primary proton output $\geq$ 1 PeV
from the central shocks.
The neutron output $\leq 0.01 $ proton primary output.\\
\\
{\large{\textbf{Acknowledgments}}}: A.Meli wishes to thank  M.Ostrowki for valuable 
comments (personal contact).

\section{\Large{References}}
{\small

Baring, M.G, \textit {$26^{th}$ ICRC}, Salt Lake City, USA, Proceedings, 1999\\
Battersby, S. J. R., Drolias, B.,  Quenby, J. J.  \textit {AstPhys} 4, 151B (1995)\\
Bednarz, J. \& Ostrowski, M. \textit {MNRAS} 283, 447-456 (1996)\\
Bednarz J.,  \textit {MNRAS} 315L.37B (2000)\\
Bell, A.R.,  \textit {MNRAS} 182,147-156 (1978)\\
Bell, A.R ,  \textit {MNRAS} 182,443-455 (1978)\\
Blanfdord, R.D $\&$ Ostriker, J.P  \textit {ApJ} 211, 793-808 (1977)\\
Ellison, D.C.,\& Jones, F.C., \& Reynolds,S.P.,  \textit {ApJ} 360, 702 (1990a)\\
Gallant, Y.A., \& Achterberg A., MNRAS 305L, 6G (1999)\\
Jones, F.C, Ellison, D.C  \textit {Sp.Sc.Rev.} 58, 259-346 1991)\\
Kirk,J.G., \& Schneider, P.,\textit{ ApJ} 315, 425 (1987a)\\
Kirk, J.G, $\&$ Schneider, P.  \textit{AJ} 322, 256-265 (1987)\\
Kirk,J.G., \& Webb, G.M.,   \textit{ApJ} 331, 336 (1988)\\
Lieu, R., \& Quenby, J.J.,  \textit {ApJ} 350, 692 (1990) \\
Meszaros,P., \& Rees, M.J.,  \textit {ApJ} 405, 278 (1993) \\
Molteni, D., Lanzafame, G., \& Chakrabarti, S. K.   \textit {ApJ}, 421, 211 (1994)\\
Ostrowski, M.,\textit {MNRAS} 249, 551-559(1991)\\
Ostrowski, M. \& Schlickeiser R.,   \textit {Sol.Ph.} 167, 381-394(1996)\\
Peacock, J.A, \textit {MNRAS} 196, 135-152(1981)\\
Quenby, J.J., \&  Lieu, R., \textit {Nature} 342, 654(1989)\\
Quenby, J.J.,  \textit {Sp.Sc.R.} 37, 201-237(1984)\\
Vietri, M.,  \textit {ApJ} 453, 883(1995)\\
Vietri, M.,  \textit {Phys.Rev.Lett.} 78, 4328V(1997)\\
Vietri, M.,  \textit {ApJLet} 488, L105(1998)\\
Waxman,E.,  \textit {Phys.Rev.Lett.} 75, 386 (1995)\\

\end{document}